\documentclass[
jcp,
reprint%
]{revtex4-1}
\usepackage{graphicx}
\usepackage{dcolumn}
\usepackage{bm}
\usepackage{color} 
\usepackage{amsmath,amssymb}
\usepackage [latin1]{inputenc}
\usepackage{epsfig}
\usepackage{bbold}
\usepackage{url}
\usepackage{subfig}
\usepackage{lineno}
\usepackage{ulem}
\usepackage{gensymb}
\usepackage{caption}
\usepackage[colorlinks=true,citecolor={blue},urlcolor={red},linkcolor={blue}]{hyperref}

\begin{document}
\title{Real-time equation-of-motion CC cumulant and CC Green's function simulations of photoemission spectra of water and water dimer}

\author{Fernando D. Vila}
\email{fdv@uw.edu}
\affiliation{Department of Physics, University of Washington, Seattle, WA, 98195, United States of America}
\author{Himadri Pathak}
\author{Bo Peng}
\author{Ajay Panyala}
\author{Erdal Mutlu}
\author{Nicholas P. Bauman}
\affiliation{Physical and Computational Sciences Directorate, Pacific Northwest National Laboratory, Richland, Washington 99354, United States of America}
\author{John J. Rehr}
\affiliation{Department of Physics, University of Washington, Seattle, WA, 98195, United States of America}
\author{Karol Kowalski}
\email{karol.kowalski@pnnl.gov}
\affiliation{Physical and Computational Sciences Directorate, Pacific Northwest National Laboratory, Richland, Washington 99354, United States of America}

\date{\today}

\begin{abstract}
Newly developed coupled-cluster (CC) methods enable simulations of ionization potentials and spectral functions of molecular systems in a wide range of energy scales ranging from core-binding to valence. This paper discusses results obtained with the real-time equation-of-motion CC cumulant approach (RT-EOM-CC), and CC Green's function (CCGF) approaches in applications to the water and water dimer molecules. We compare the ionization potentials obtained with these methods for the valence region with the results obtained with the CCSD(T) formulation as a difference of energies for $N$ and $N-1$ electron systems.
All methods show good agreement with each other. They also agree well with experiment, with errors usually below 0.1 eV for the ionization potentials.
We also analyze unique features of the spectral functions, associated with the position of satellite peaks, obtained with the RT-EOM-CC and CCGF methods employing single and double excitations, as a function of the monomer OH bond length and the proton transfer coordinate in the dimer. Finally, we analyze the impact of the basis set effects on the quality of calculated ionization potentials and find that the basis set effects are less pronounced for the augmented-type sets.
\end{abstract}

\maketitle

\section{Introduction} 
\label{intro}
Coupled-cluster (CC) theory \cite{coester58_421,coester60_477,cizek66_4256,paldus72_50,purvis82_1910,paldus1999critical,RevModPhys.79.291}
has evolved into one of the most accurate formulations to capture complex correlation effects defining a broad class of quantum effects that drive chemical transformations, usually identified with bond-forming and bond-breaking processes.\cite{RevModPhys.79.291,Bartlett2009} Over the last few decades, theoretical efforts closely followed by the development of sophisticated computational models that take advantage of the ever-growing computational power of parallel architectures, enabled the establishment of a hierarchy of approximations and novel formulations for closed/open-shell molecules, strong correlation, and large-scale systems. Special attention has been paid to single-reference CC formulations that can adequately describe chemical reactions and topological events of the corresponding ground-state potential energy surfaces, including barriers, avoided crossings, and multiple minima. This effort includes an impressive array of formulations based on the inclusion of high-rank collective effects contributing to the expansion of the ground-state wave function. This progress was made possible to achieve thanks to cornerstone implementations of the now ubiquitous models such as CCSD,\cite{purvis82_1910} CCSDT,
\cite{ccsdt_noga,ccsdt_noga_err,scuseria_ccsdt}
CCSDTQ \cite{Kucharski1991,ccsdtq_nevin} and their perturbative counterparts, including CCSD[T],\cite{urban1985towards,noga1987towards} CCSD(T),\cite{raghavachari89_479} CCSD(TQ) and CCSDT(Q) type approaches.
\cite{bartlett1990non,kucharski1998efficient}

For its simplicity, the CCSD(T) method has assumed a unique position in high-accuracy chemical simulations of equilibrium properties and chemical reactions.
However, a significant effort has been expended to alleviate serious issues associated with the perturbative nature of the (T) for molecular configurations away from the equilibrium geometry.\cite{crawford_t,mmcc1,crccx,deustua2017converging,gwal1,gwal2,sohir1,ybom1,robkn,ugur1,mauss1}
Different design principles drive a large class of these developments in perturbative and non-perturbative formulations.

Structural changes in molecular systems such as bond-breaking/forming can be characterized by analyzing their potential energy surface (PES) or scrutinizing processes/states in various energy regimes sensitive to the geometry changes induced by chemical transformations. Excited state extensions of CC formalism play an essential role in these studies either through equation-of-motion CC methodologies (EOMCC) 
\cite{bartlett89_57,bartlett93_414,Stanton1993} or CC linear response theory (LR-CC),\cite{monkhorst77_421,Koch1990} which are usually formulated in frequency space. These methods have been widely applied in studies of excitation energies, excited-state PESs, excited-state non-adiabatic dynamics, frequency-dependent optical properties, and, more recently, multi-component polaritonic systems.\cite{PhysRevX.10.041043} Green's function extensions of the CC formalism (CCGF) 
\cite{nooijen92_55,nooijen93_15,nooijen95_1681,meissner93_67,kkgfcc1,kkgfcc2,kkgfcc3,peng2021coupled,mcclain2016spectral,zhu2019coupled,shee2019coupled,lange2018relation}
can also capture excited-state correlation effects needed to describe quasiparticles (QP) and satellite peaks observed in X-ray photoemission spectra (XPS). 
It is worth mentioning that the development of the CC methods for the description of satellite peaks was paralleled by enabling  vertex corrected GW$\Gamma$ approaches \cite{mejuto2021multi} and adaptive sampling configuration interaction (ASCI) algorithms
\cite{tubman2020modern} capable of attaining high accuracy in describing multi-configurational electronic states for relatively little computational cost. 
 
Time-dependent CC formulations, which provide an alternative way of describing excited-state processes, have attracted significant attention in the last decade. \cite{kvaal2012ab,pedersen2019symplectic,sato2018communication,nascimento2016linear,nascimento2017simulation,cooper2021short,li2020real}
In this context, we have recently developed 
a real-time equation-of-motion CC (RT-EOM-CC) method\cite{RVKNKP,vila2021eom,vila2020real,vila2022real} to compute the one-electron GF based on a CCSD cumulant approach, which provides several advantages compared to other excited-state formulations.
For example, the approach leads to an explicit exponential cumulant representation of the Green's function\cite{Hedin99review} and, at the same time, allows one to formulate a non-perturbative expression for the cumulant in terms of the solutions to a set of coupled, first-order nonlinear differential equations for the CC amplitudes. Additionally, the RT-EOM-CC formulation provides a theoretical platform for including nonlinear corrections to the traditional cumulant approximations, which are usually linear in the self-energy. 

In this paper, we compare the performance of the CCGF and RT-EOM-CC formulations when applied to core and valence binding energies of the water and water dimer molecules. In particular, we focus our analysis on identifying unique features of spectral functions, i.e., the location of quasiparticle and satellite peaks (for core and valence regions) as functions of geometry changes in bond-breaking processes. In this regard, we focus on the O-H stretch in H$_2$O and the bridging proton transfer process in (H$_2$O)$_2$.  We demonstrate that the impact of the geometry changes is especially strong for the satellite region of the core spectral function. We also investigate  the effect of basis set effects on the accuracies of ionization potentials obtained with the CCGF and RT-EOM-CC models with singles and doubles (CCSDGF/RT-EOM-CCSD) and compare them with large CCSD(T) calculations.

The paper is organized as follows: 
In Sec.~\ref{theory}, we provide a brief overview of   the CCGF, IP-EOMCC, and RT-EOM-CC formulations and  approximations used in the calculations for the H$_2$O and (H$_2$O)$_2$ systems. In Sec.~\ref{results}, we discuss the ionization potentials and spectral functions in the valence and core regions. Using RT-EOM-CCSD we investigate how bond-breaking process affects the spectral functions calculated for core and valence energy regimes. Special focus is on evaluating the performance of the RT-EOM-CCSD formalism in comparison with other CC formulations. In Sec.~\ref{results} we also discuss the basis set effects on the accuracy of calculated ionization potentials.  Finally, Sec.~\ref{conclusion} briefly summarizes our results.

\section{Theory}
\label{theory}
In this section, we discuss basic aspects of the various CC approaches used here to calculate correlated binding energies in various energy windows. In particular, we focus on the coupled-cluster Green's function formalism and closely related ionization-potential equation-of-motion coupled-cluster (EOM-CC) approach, and on the real-time EOM-CC cumulant formulation. 

\subsection{CC Green's function}

The retarded part of the Green's function (advanced part can be developed in an analogous way) is defined by the following matrix elements $G^R_{pq}(\omega)$:
\begin{equation}
G^R_{pq}(\omega) =
\langle \Psi_g | a_q^\dagger (\omega + ( H - E_0 ) - i \eta)^{-1} a_p | \Psi_g \rangle \;,
\label{gf0}
\end{equation}
where $E_0$ is the corresponding ground-state energy for the $N$-electron system, $\eta$ is a broadening factor, 
$a_p/a^{\dagger}_p$ are creation/annihilation operators for an electron in the $p$-th spin-orbital,
and $|\Psi_g\rangle$ is the ground state wave function of the $N$-electron system. The bi-variational CC
\cite{arponen1983variational,Stanton1993} formalism  utilizes  distinct parametrizations for the bra ($\langle\Psi_g|$) and ket 
($|\Psi_g\rangle$) ground-state wave functions,
i.e., 
\begin{eqnarray}
\langle \Psi_g | &=& \langle \Phi | (1+\Lambda)e^{-T}  \label{biv1} \\
| \Psi_g \rangle &=& e^T | \Phi \rangle,
\label{biv2}
\end{eqnarray}
which leads to the following form of retarded part of the CC Green's function (CCGF) 
\cite{nooijen92_55,nooijen93_15,nooijen95_1681,meissner93_67,kkgfcc1,mcclain2016spectral,kkgfcc3,
PengKowalski2018,doi:10.1063/1.5138658,doi:10.1021/acs.jctc.9b00172}
\begin{eqnarray}
G^R_{pq}(\omega) = 
\langle\Phi|(1+\Lambda) \overline{a_q^{\dagger}} (\omega+\bar{H}_N- \text{i} \eta)^{-1} 
	\overline{a_p} |\Phi\rangle. 
\label{gf1}
\end{eqnarray}
The similarity transformed operators $\overline{A}$ ($A = H, a_p, a_q^{\dagger}$) are defined as $\overline{A} = e^{-T} A ~e^{T}$ (the $\overline{H}_N$ stands for a normal product form of $\overline{H}$). The numerical algorithms for calculating Eq. (\ref{gf1})
employ  auxiliary  operators $X_p(\omega)$
\begin{eqnarray}
\hspace*{-0.4cm} X_p(\omega) 
&=& X_{p,1}(\omega)+X_{p,2}(\omega) + \ldots \notag \\
&=& \sum_{i} x_i(\omega)_p  a_i  + \sum_{i<j,a} x_{ij}^a(\omega)_p a_a^{\dagger} a_j a_i +\ldots ,
\;\; \forall_p \label{xp} 
\end{eqnarray}
that satisfy the equations 
\begin{eqnarray}
(\omega+\overline{H}_N - \text{i} \eta )X_p(\omega)|\Phi\rangle = 
	\overline{a_p} |\Phi\rangle.  \label{eq:xplin} 
\end{eqnarray}
In Eq. (\ref{xp}), the $i,j,\ldots$ ($a,\ldots$) labels refer to spin-orbitals labels occupied (unoccupied) in the $N$-electron reference function $|\Phi\rangle$.
Using these operators, the matrix elements 
can be expressed in a simple form 
\begin{eqnarray}
G^R_{pq}(\omega) = 
\langle\Phi_g|(1+\Lambda) \overline{a_q^{\dagger}} X_p(\omega) |\Phi_g\rangle.
\label{gfxn2}
\end{eqnarray}
The tensors $x_i(\omega)_p$,  $x_{ij}^a(\omega)_p$, etc. in Eq.~(\ref{xp}) represent sought after cluster amplitudes.
In our implementation of CCGF, we approximate the 
de-excitation $\Lambda$ operator by the cluster operator $T^{\dagger}$.
The main numerical effort associated with constructing the retarded CC Green's function is associated with the need to solve a large number of independent linear equations, which in turn can contribute to efficient parallel schemes utilizing multiple levels of parallelism. 
We implemented the CCGF model with single and double excitations (CCSDGF)\cite{peng2021GFCCLib} using the parallel tensor library TAMM (Tensor Contraction for Many-body Methods).\cite{TAMM} The current version of the CCGF code utilizes Cholesky-decomposed two-electron integrals that significantly reduce the memory requirements and inter-node communication.

\subsection{Ionization potential EOMCC formalism}

The ionization potential EOMCC (IP-EOMCC) formulation \cite{stanton1995perturbative}
utilizes the EOMCC type wave function expansion for the $k$-th electronic state $|\Psi_k\rangle$ of the $N-1$ electron system:
\begin{equation}
    |\Psi_k\rangle=R_k e^T |\Phi\rangle,
    \label{ipcc1}
\end{equation}
where $|\Phi\rangle$ and $T$ are the reference function and cluster operator of the $N$-electron system, respectively, and the $R_k$ operator is given by the expansion 
\begin{equation}
    R_k=\sum_i r_i(k)a_i + \sum_{i<j;a} r(k)_{ij}^a a_a^{\dagger} a_j a_i + \ldots ~~ .
    \label{ipcc2}
\end{equation}
The $R_k$ can be determined by solving a non-Hermitian eigenvalue problem in the $N-1$ Hilbert space
\begin{equation}
    \bar{H}_N R_k |\Phi\rangle = \omega^{\rm IP}_k 
    R_k |\Phi\rangle \;,
    \label{ipcc3}
\end{equation}
where $\omega_k^{\rm IP}$ stands for the $k$-th ionization potential. It should be noted that the eigenvalues of the IP-EOMCC Eq. (\ref{ipcc3}) are identical to the singularities (as functions of $\omega$ parameter for $\eta=0$) of the solutions of Eq. (\ref{eq:xplin}).
For the present studies we use the TCE implementation of the IP-EOMCC model with singles and doubles (IP-EOMCCSD).\cite{bhaskaran2014equation}

\subsection{Real-time EOM-CC cumulant approach}

The cumulant approximation, in which the core-hole GF is approximated as the product of the free particle GF and the exponential of a cumulant, has been shown to provide accurate spectral functions for extended systems.\cite{PhysRevB.90.085112,PhysRevB.91.121112,PhysRevB.94.035156,doi:10.1116/6.0001173} We have recently combined this form of the GF with the CC method to develop\cite{doi:10.1063/5.0004865,vila2021eom,vila2020real,vila2022real} a real-time equation-of-motion CC (RT-EOM-CC) approach to compute the core and valence one-electron GF based on a CC approximation to the cumulant. In this approximation, the retarded GF can be expressed as
\begin{equation}
\label{eq:cum_gf}
G_{c}^{R}(t) = -i \Theta(t) e^{-i (\epsilon_c + E_N^{corr}) t} e^{C_c^{R}(t)}.
\end{equation}
where $c$ is the index of the excited hole spin-orbital (which in our previous studies was restricted to be a core spin-orbital, but here can be any occupied spin-orbital), $\epsilon_c$ is its bare or Fock energy, $E_N^{corr}$ is the correlation energy of the $N$ electron ground state, and $C_c^{R}(t)$ is the retarded cumulant associated with $c$.
As shown in Eq. \ref{eq:cum_gf}, the CC approximation naturally results in a GF with an explicit exponential cumulant form,\cite{langreth70,SG1978,Hedin99review,sky} which, unlike that obtained in self-energy formulations, is non-perturbative. The expression for the time derivative of the cumulant in terms of the time-dependent CC amplitudes takes the simple form:
\begin{equation}
\label{eq:dcdt}
\begin{split}
-i\frac{d C_c^R(t)}{dt} =& \sum_{ia} f_{ia} t_i^a(t)
 + \frac{1}{2} \sum_{ijab} v_{ij}^{ab} t_j^b(t) t_i^a(t)\\
 +& \frac{1}{4} \sum_{ijab} v_{ij}^{ab} t_{ij}^{ab}(t),
\end{split}
\end{equation}
where the $f_{pq} = \epsilon_p \delta_{pq} - v_{pc}^{qc}$ are the matrix elements of the $N-1$ electron Fock operator, $\epsilon_p$ is the energy of spin-orbital $p$, and $v_{pq}^{rs} = \left< pq \left| \right| rs \right>$ are the antisymmetrized two-particle Coulomb integrals over the generic spin-orbitals $p, q, r, s$. The time-dependent amplitudes $t_{ij...}^{ab...}(t)$ of the CC expansion are obtained from a set of coupled, first-order non-linear differential equations analogous to those used in the solution of the static CCSD equations (see, for instance, Ref. \citenum{vila2022real}). These equations have the initial conditions $t_{ij...}^{ab...}(0)=0$, which result in the initial condition $C_c^R(0)=0$ for the cumulant in Eq \ref{eq:cum_gf}.
As shown in Eq. \ref{eq:dcdt}, the RT-EOM-CC cumulant also naturally includes non-linear contributions, in contrast to traditional cumulant approximations that usually depend linearly on the self-energy. RT-EOM-CC uses two main approximations: 1) The separable approximation
$\left| 0 \right> \simeq a_c^\dagger \left| N-1 \right>$.
Here $\left| N-1 \right>$ is the fully correlated $N-1$ electron
portion of the $N$ electron exact ground state wave function $\left| 0 \right>$. 2) A time-dependent (TD) CC ansatz for this $N-1$ electron state, i.e., $\left| N-1, t \right> = \tilde N(t) e^{T(t)} \left| \phi \right>$. Here $\tilde N(t)$ is the normalization factor and $T(t)$ is the TD cluster operator. In contrast with traditional CC formulations, $T(t)$ is defined in the $N-1$ electron Fock space. Therefore, the reference determinant $\left| \phi \right>$ has a hole in level $c$, i.e., $\left| \phi \right> = a_c \left| \Phi \right>$, where $\left| \Phi \right>$ is the traditional ground state $N$ electron Hartree--Fock (HF) Slater determinant. We have shown\cite{vila2022real} that, at the CCSD level, the RT-EOM-CC method gives accurate core quasiparticle (QP) binding energies, with a mean absolute error (MAE) from experiment of about 0.3 eV for the CH$_4$, NH$_3$, H$_2$O, HF and Ne systems. The method also gives a good treatment of the many-body satellites, with errors in the QP-satellite gap of less than 1 eV. Finally, despite the use of the separable approximation, RT-EOM-CCSD provides accurate valence ionization energies, as shown below and elsewhere.\cite{vilartvalence2022}

\section{Results and Discussion}
\label{results}

\subsection{Geometries, basis sets, and computational details}

The calculations of the water monomer use experimental geometries\cite{cccbdb} where R(OH) = 0.958 \AA\ and $\theta$(HOH)= 104.48\degree. The molecule has $C_{2v}$ symmetry and is oriented in the traditional way, with the atoms lying in the $yz$ plane. This results in the molecular orbital configuration $(1$a$_1)^2(2$a$_1)^2(1$b$_2)^2(3$a$_1)^2(1$b$_1)^2$. For the water dimer structure we use the best estimate reported by Lane\cite{lane2013ccsdtq} which was obtained using the counterpoise corrected CCSD(T)-F12b method at the complete basis set limit, with corrections for higher-order excitations, core correlation, relativistic effects, and diagonal Born-Oppenheimer effects. For reference, this structure has the following parameters: R(OH)$_f$ = 0.95685 \AA, R(OH)$_b$ = 0.96414 \AA, R(OH)$_a$ = 0.95843 \AA, $\theta$(HOH)$_d$ = 104.854\degree, $\theta$(HOH)$_a$ = 104.945\degree, R(O$\cdots$O) = 2.90916 \AA, $\alpha$ = 5.686\degree, $\beta$ = 123.458\degree (for an explanation of these labels refer to Fig. 3 of Ref. \citenum{lane2013ccsdtq}). The proton transfer coordinate $\Delta$R(OH)$_b$ used here is defined as the otherwise rigid displacement of the bridging proton, around its optimal position, following the direction parallel to the axis defined by the O$\cdots$O bond.

The RT-EOM-CCSD simulations were performed using the Tensor Contraction Engine (TCE) \cite{TCE2} implementation of RT-EOM-CCSD.\cite{vila2022real} All the simulations used a time step of 0.025 au and frozen virtual orbitals above 10 au, when present. The total length of the simulations was 90 au for the valence states of the monomer and dimer, and 400 au for the core state of the monomer. The valence simulations of the monomer used the aug-cc-pVTZ basis set,\cite{augccpvdz} while the core ones used the Sapporo-TZP\cite{sapporotzp} basis set. The RT-EOM-CC simulations of the dimer used the aug-cc-pVTZ basis set.

The CCGF spectral functions were calculated using the highly-scalable parallel tensor library TAMM (Tensor Algebra for Many-body Methods) \cite{mutlu2022tamm, mutlu2019toward} implementation of the CCSDGF formalism.\cite{peng2021GFCCLib, doi:10.1063/1.5138658}
The valence vertical ionization potentials for the water and water dimer were calculated using IP-EOMCCSD code available in NWChem.\cite{nwchem, apra2020nwchem}

All CCSD(T) calculations for the water dimer in cc-pVXZ and aug-cc-pVXZ (X=D, T, Q, 5, 6) were obtained with the TAMM implementation of the CCSD(T) formalism,\cite{kim2020scalable,mutlu2022tamm, mutlu2019toward} with a linear dependence threshold for the basis sets of 10$^{-5}$, SCF convergence cutoff of 10$^{-8}$ for the energy, and a CCSD convergence cutoff of 10$^{-8}$. In the largest CCSD(T) calculations corresponding to the aug-cc-pV6Z, more than 1,100 orbitals were correlated.

\begin{center}
\begin{table}[t]
\centering
\caption{The low-lying ionization potentials of the H$_2$O molecule in experimental geometry for various basis sets obtained with the IP-EOMCCSD, relativistic IP-EOMCCSD, and RT-EOM-CCSD approaches.}
\begin{tabular}{lccc}
\hline
\hline\\
Basis set   &\, $1^2B_1$ &\, $1^2A_1$ &\, $1^2B_2$ \\
\hline\\
            & \multicolumn{3}{c}{IP-EOMCCSD}  \\
\hline
cc-pVDZ     &\, 11.807 &\, 14.131 &\, 18.470  \\
cc-pVTZ     &\, 12.423 &\, 14.651 &\, 18.848  \\
cc-pVQZ     &\, 12.707 &\, 14.912 &\, 19.072  \\
cc-pV5Z     &\, 12.721 &\, 14.920 &\, 19.079  \\
aug-cc-pVDZ &\, 12.384 &\, 14.673 &\, 18.906  \\
aug-cc-pVTZ &\, 12.617 &\, 14.832 &\, 19.002  \\
aug-cc-pVQZ &\, 12.707 &\, 14.912 &\, 19.072  \\
aug-cc-pV5Z &\, 12.740 &\, 14.938 &\, 19.095  \\
\hline \\
 & \multicolumn{3}{c}{Relativistic IP-EOMCCSD}  \\
\hline
 dyall.ae4z &\, 12.627 &\, 14.834 &\, 18.871  \\
 \hline \\
            & \multicolumn{3}{c}{RT-EOM-CCSD}  \\
\hline
aug-cc-pVDZ  &\, 12.477 &\, 14.774 &\, 18.981  \\
aug-cc-pVTZ  &\, 12.587 &\, 14.817 &\, 18.977  \\
\hline \\
  & \multicolumn{3}{c}{Expt.}  \\
\hline
Ref. \citenum{h2oIPnist}  &\,  12.621\,$\pm$\,0.002   &\,                    &\,  \\
Ref. \citenum{doi:10.1021/jp030263q} &\, &\, 14.84\,$\pm\,$0.02 &\, 18.78\,$\pm$\,0.02 \\
\hline
\hline 
\end{tabular}
\label{table_h2o_ipeom}
\end{table}
\end{center}

\begin{figure}[t]
	\includegraphics[clip,trim=0.5cm 1.0cm 0.5cm 2.8cm,
	width=0.50\textwidth]{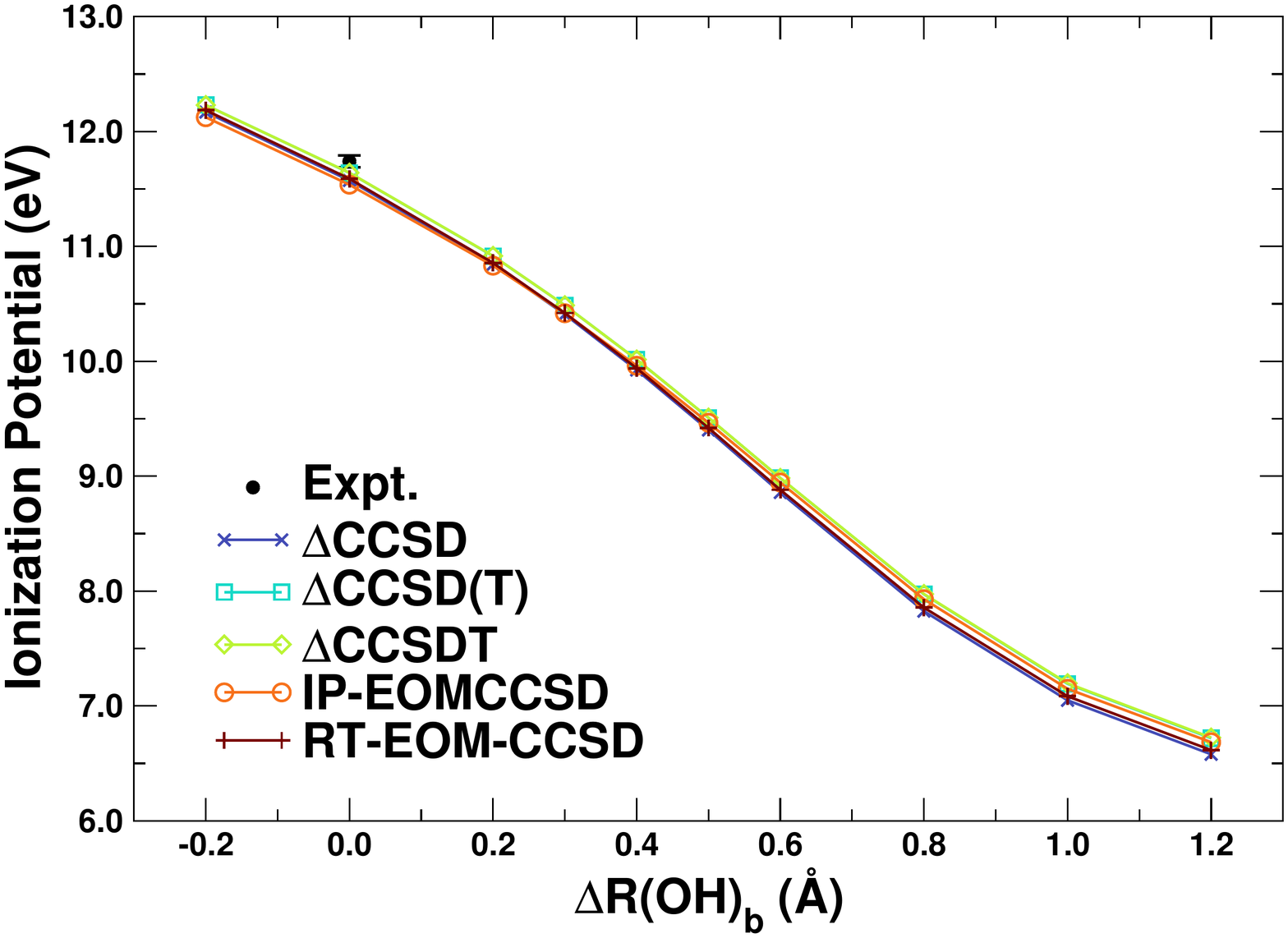}
	\includegraphics[clip,trim=0.5cm 1.0cm 0.5cm 2.8cm,
	width=0.50\textwidth]{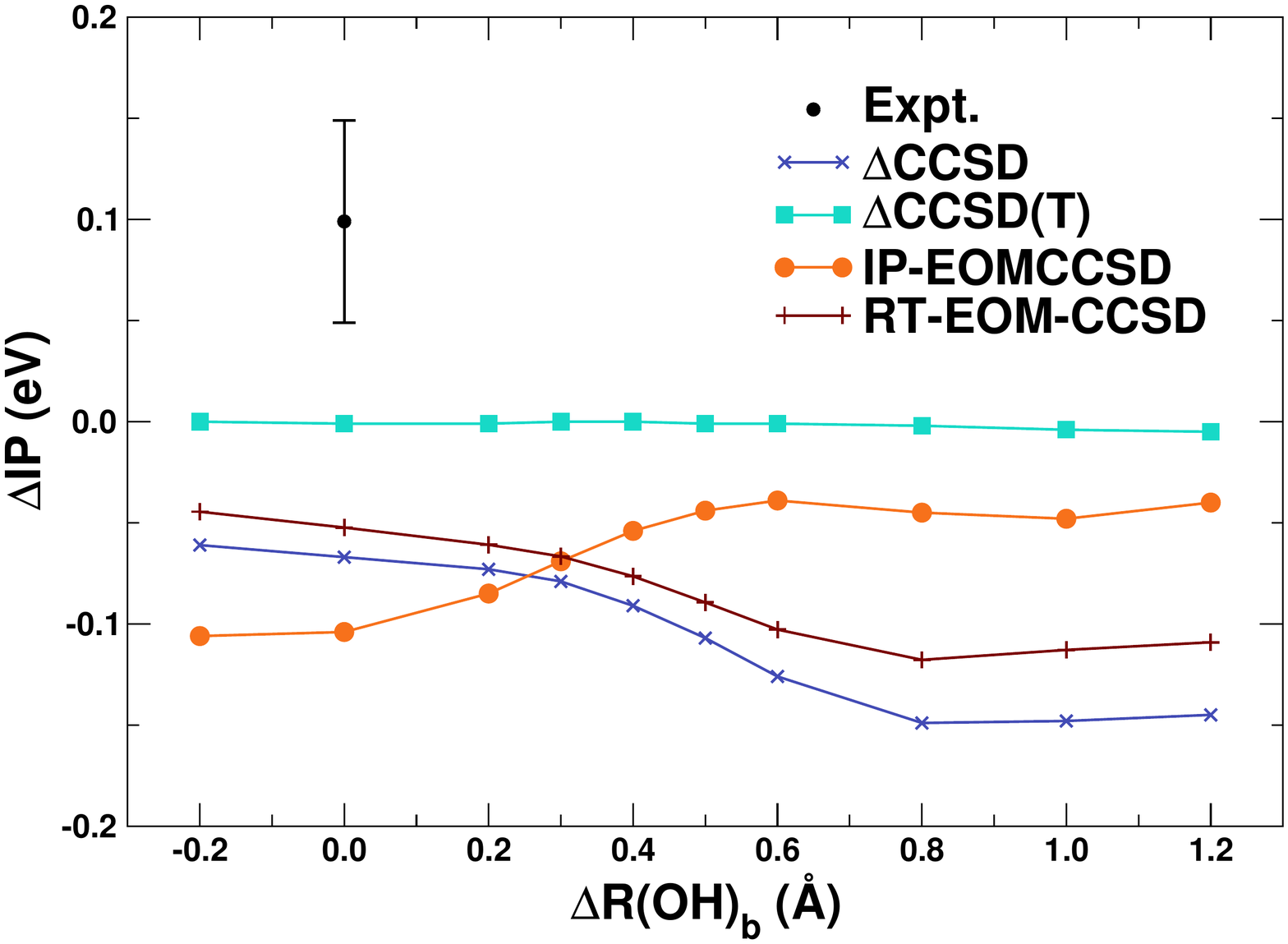}
	\caption{Top: Ionization potential of the $1^2A''$ state of (H$_2$O)$_2$ as a function of proton transfer coordinate calculated with the IP-EOMCCSD and RT-EOM-CCSD approaches, and with the $\Delta$CCSD, $\Delta$CCSD(T) and $\Delta$CCSDT methods as energy differences between the $N$ and $N-1$ electron systems. All calculations used the the aug-cc-pVDZ basis and correlated all the molecular.Bottom: Ionization potentials relative to the $\Delta$CCSDT values.}
\label{fig:methcomp}
\end{figure}

\subsection{Valence region}

Table \ref{table_h2o_ipeom} collects the IP-EOMCCSD, relativistic IP-EOMCCSD, and RT-EOM-CCSD results for the ionization potentials corresponding to the $1^2B_1$, $1^2B_2$ and $1^2A_1$ states of H$_2$O$^+$ (in the equilibrium structure of H$_2$O) calculated with various basis sets of the cc-pVXZ \cite{ccpvdz}and aug-cc-pVXZ (X=D,T,Q,5) \cite{augccpvdz} type. The observed basis set effects can vary with the targeted states. For example, the difference between cc-pVDZ and cc-pV5Z corresponding to the lowest ionization potential ($1^2B_1$) amounts to 0.917 eV. The analogous difference for the $1^2B_2$ state is equal to 0.609 eV. The utilization of larger aug-cc-pVXZ basis sets significantly reduces the dependence of the calculated IP-EOMCCSD ionization energies on the basis set employed, which is well illustrated by the $1^2B_1$ ionization potential, where the difference between IP-EOMCCSD results for aug-cc-pVDZ and aug-cc-pV5Z basis sets is reduced to 0.356 eV.
It should also be noted that increasing the basis set size in the IP-EOMCCSD simulations of the $1^2B_1$ state results in overshooting its experimental value (12.621$\pm$0.002). The aug-cc-pVTZ basis set provides the best IP-EOMCCSD estimate for this state. In analogy to the IP-EOMCCSD method, the RT-EOM-CCSD simulations using the same aug-cc-pVTZ basis set provide an accurate (12.587 eV) estimate of the experimental value. The RT-EOM-CCSD results also agree with the IP-EOMCCSD results and with experiment for the aug-cc-pVTZ basis set for the remaining states.
Table \ref{table_h2o_ipeom} also shows results obtained with the fully 4-component relativistic IP-EOMCCSD.\cite{pathak2014atom,pathak2014molecule,pathak2016open} The effects of relativity have been included by using a Dirac--Coulomb--Gaunt Hamiltonian, taken a Gaussian charge density distribution nuclear model to mimic the finite nucleus. The calculations use the dyall.ae4z basis for both H and O, where large and small component basis functions are uncontracted. All electrons are correlated, and none of the virtual spinors are frozen. The computed IPs are found to be very precise in comparison to experiment. However, the effect of relativity has a negligible role.

\renewcommand{\tabcolsep}{0.55cm}
\begin{center}
\begin{table}
\centering
\caption{Low-lying ionization potentials of (H$_2$O)$_2$ in the equilibrium geometry, obtained with the IP-EOMCCSD and RT-EOM-CCSD approaches for various  basis sets. In all calculations all molecular orbitals were correlated.}
\begin{tabular}{lcc}
\hline
\hline\\
Basis set &\, $1^2A''$ &\, $1^2A'$  \\
\hline \\
 & \multicolumn{2}{c}{IP-EOMCCSD}  \\
\hline
cc-pVDZ       &\,  10.889  &\,  12.369 \\
cc-pVTZ       &\,  11.546  &\,  12.938 \\
cc-pVQZ       &\,  11.774  &\,  13.136 \\
cc-pV5Z       &\,  11.868  &\,  13.216 \\
aug-cc-pVDZ   &\,  11.537  &\,  12.873 \\
aug-cc-pVTZ   &\,  11.768  &\,  13.107 \\
aug-cc-pVQZ   &\,  11.857  &\,  13.200 \\
\hline \\
 & \multicolumn{2}{c}{RT-EOM-CCSD}  \\
\hline 
aug-cc-pVDZ   &\,  11.589  &\,  13.043 \\
\hline \\
 & \multicolumn{2}{c}{Expt.}  \\
\hline 
Ref. \citenum{tomoda1982photoelectron}   &\, 12.1\,$\pm$\,0.1 &\, 13.2\,$\pm$\,0.2   \\
Ref. \citenum{stephen_leone_water_dimer} &\, 11.74\,$\pm$\,0.05 &\,  \\
\hline
\hline 
\end{tabular}
\label{table_h2o2_ipeomccsd_b}
 \end{table}
\end{center}

Fig.~\ref{fig:methcomp} (top) shows a comparison of the ionization potential of the $1^2A''$ state of (H$_2$O)$_2$ as a function of proton transfer coordinate calculated with the IP-EOMCCSD and RT-EOM-CCSD approaches, and with the $\Delta$CCSD, $\Delta$CCSD(T) and $\Delta$CCSDT methods as energy differences between the $N$ and $N-1$ electron systems. Fig. \ref{fig:methcomp} (bottom) shows the same ionization potentials relative to those obtained with the  $\Delta$CCSDT method.
For all geometries considered, the $\Delta {\rm CCSD(T)}$ results are in excellent agreement with the $\Delta {\rm CCSDT}$ ones. At the same time $\Delta {\rm CCSD(T)}$ results provide systematic improvements of the $\Delta {\rm CCSD}$ ionization potentials. It is also worth noticing that the IP-EOMCCSD results are consistently  better for the larger separations (using $\Delta {\rm CCSDT}$ results as a reference point) than the $\Delta {\rm CCSD}$ ones.

The effect of the basis set quality on the low-lying ionization energies calculated with the IP-EOMCCSD formalism for the equilibrium structure of the water dimer is shown in Table \ref{table_h2o2_ipeomccsd_b}. As in the case of the water monomer (see Table~\ref{table_h2o_ipeom}), one can observe a significant impact of the basis set size on the calculated values of the ionization energies. In particular, the same trend of increasing the IP values with the growing size of the basis set can be noticed. 
For example, for the aug-cc-pVXZ (X=D,T,Q) the aug-cc-pVQZ basis is needed to get a satisfactory agreement with the experiment for the IP corresponding to the $1^2A''$ state. 
In the case of the $1^2A'$ state, the IP-EOMCCSD results align with the experimental values. The RT-EOM-CCSD result in the aug-cc-pVDZ basis  for the $1^2A''$ 
are better than the IP-EOMCCSD ionization potential obtained with the same basis set.


\begin{table}[t]
\caption{\label{tab:gs}{Ionization potential of the $1^2A''$ state of the water dimer as a function of basis set, computed as the energy difference between the CCSD(T) ground state energies of (H$_2$O)$_2$ and (H$_2$O)$_2^+$.}}
\begin{ruledtabular}
\begin{tabular}{lcc}
\multicolumn{1}{l}{Basis set}&\multicolumn{1}{c}{\# of basis functions}& \multicolumn{1}{r}{IP (eV)}\\
           
\hline
cc-pVDZ &\, 50  &\,   10.998           \\
cc-pVTZ &\, 130  &\,  11.596              \\
cc-pVQZ &\, 280  &\,  11.787               \\
cc-pV5Z &\, 525 &\,  11.865              \\
cc-pV6Z &\, 848 &\,     11.886           \\
\hline
aug-cc-pVDZ &\, 86 &\,  11.643               \\
aug-cc-pVTZ &\, 210 &\, 11.804      \\
aug-cc-pVQZ &\, 428  &\, 11.862      \\
aug-cc-pV5Z &\, 741   &\, 11.885   \\
aug-cc-pV6Z &\, 1120 &\, 11.893     \\
\end{tabular}
\end{ruledtabular}
\label{ccsd_t}
\end{table}

In order to assess the basis set size effects on the ionization potentials, we have performed a series of calculations for the lowest ionization potential of the water dimer changes. These results are compiled in Table \Ref{ccsd_t}. As described above, the ionization potential is calculated as the difference in energy between the CCSD(T) ground state energies of the water dimer and its cation at the optimal geometry. The basis set sizes vary from the smallest set (cc-pVDZ), with 50 basis functions for the water dimer, to the largest (aug-cc-PV6Z) with 1120 basis functions. Interestingly, the augmentation functions seem to significantly improve the convergence of the ionization potential to its basis set limit, as demonstrated by the smaller (0.25 eV) range of IP values for the augmented sets than for the unaugmented ones (1 eV). From the table, it is evident that even a modest augmented basis set like aug-cc-pVDZ does a job compared to a much more saturated basis set.


\begin{figure}[t]
	\includegraphics[clip,trim=0.5cm 1.0cm 0.5cm 2.8cm,
	width=0.45\textwidth]{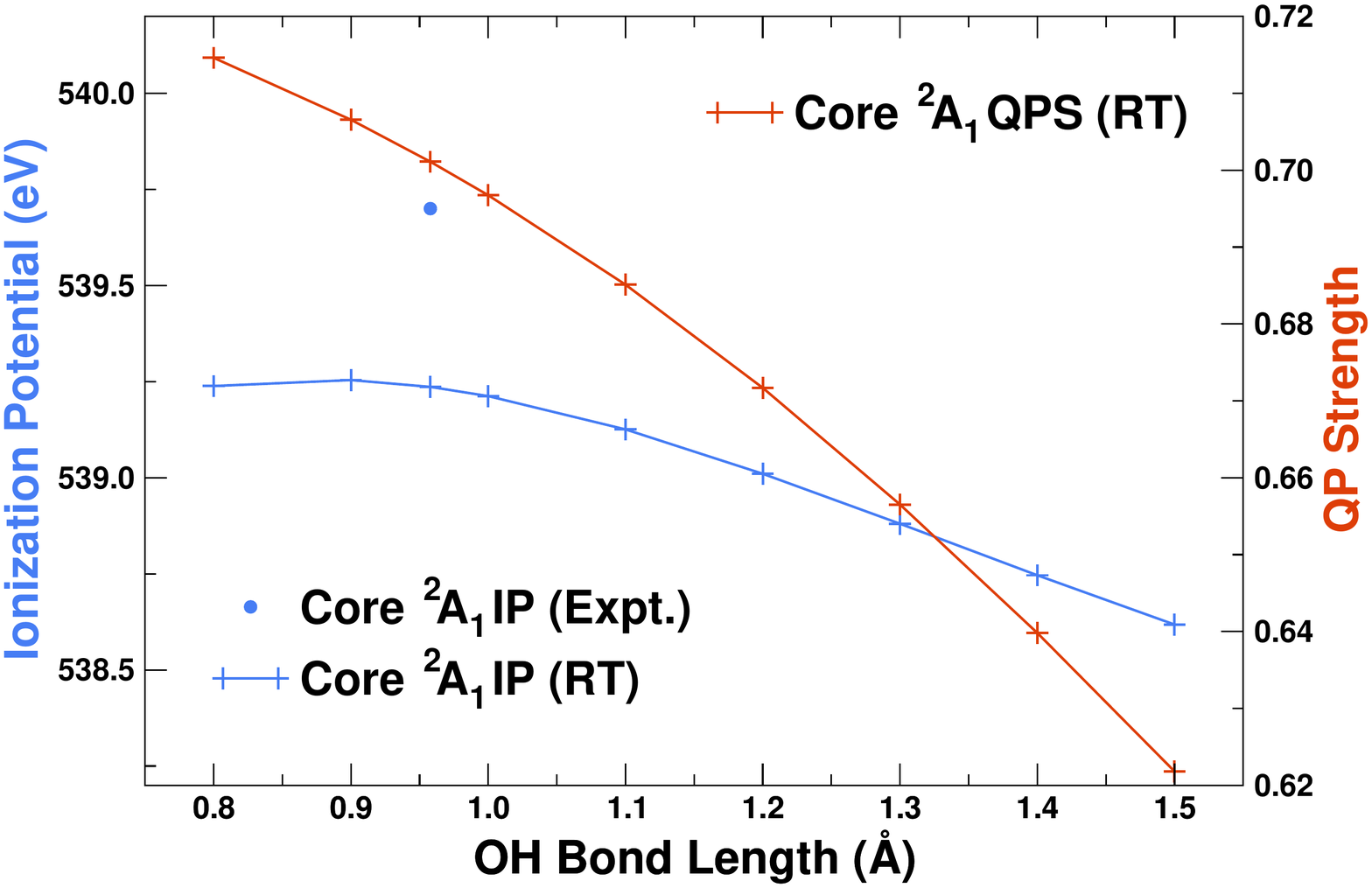}
	\includegraphics[clip,trim=0.5cm 1.0cm 0.5cm 2.8cm,
	width=0.45\textwidth]{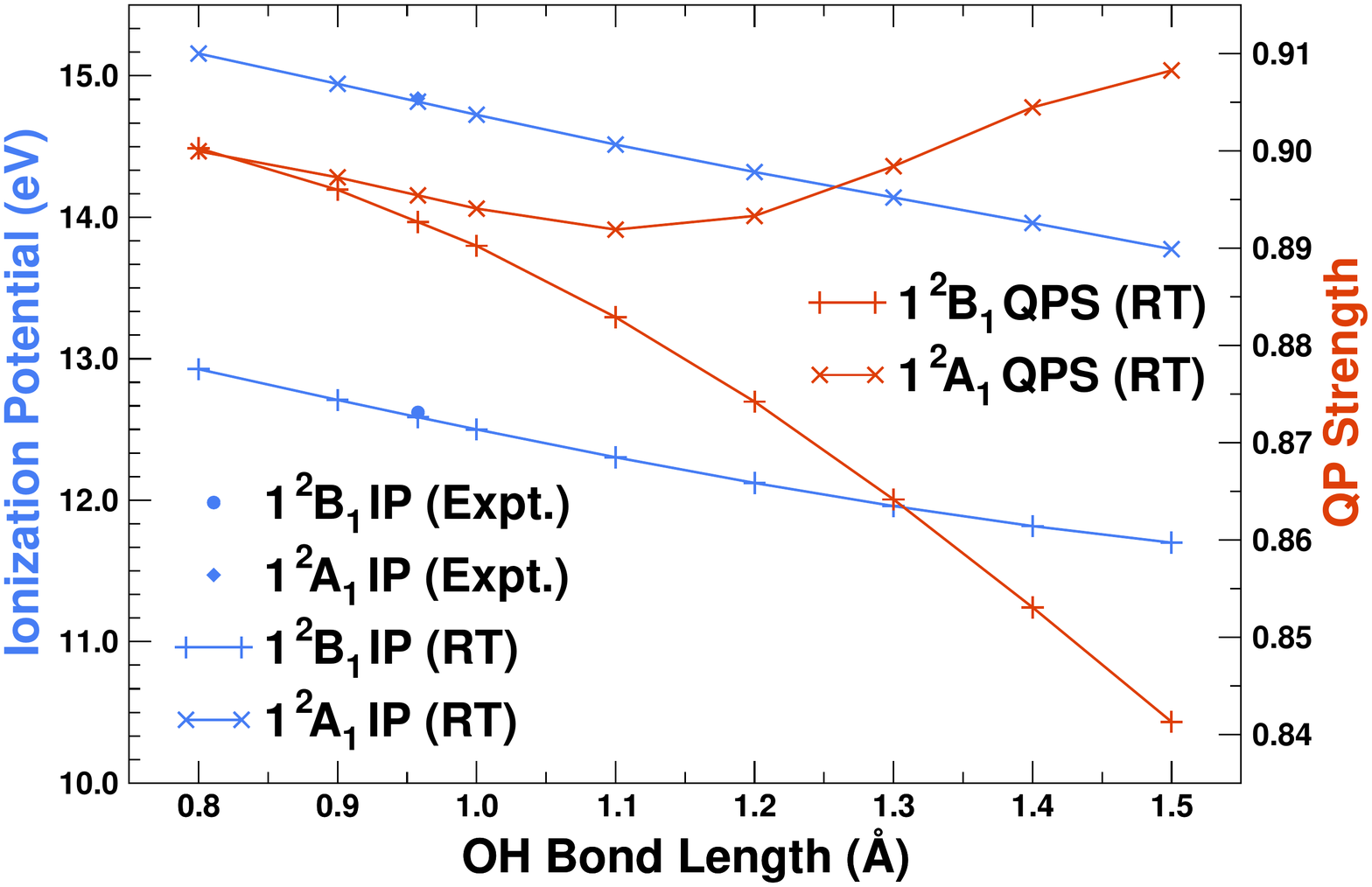}
	\caption{Ionization potential (IP) and quasi-particle strength (QPS) as a function of OH bond length for the core $^2A_1$ (top), and HOMO-1 $1^2A_1$ and HOMO $1^2B_1$ (bottom) states of H$_2$O, calculated with the RT-EOM-CCSD method (RT) using the aug-cc-pVTZ basis set for the valence states and the Sapporo-TZP\cite{sapporotzp} for the core. The blue circles and diamonds indicate the experimental (Expt.) ionization energies at equilibrium.\cite{h2ospf,doi:10.1021/jp030263q}}
\label{fig:h2oIP}
\end{figure}

\begin{figure}
	\includegraphics[clip,trim=0.5cm 1.0cm 0.5cm 2.8cm,
	width=0.45\textwidth]{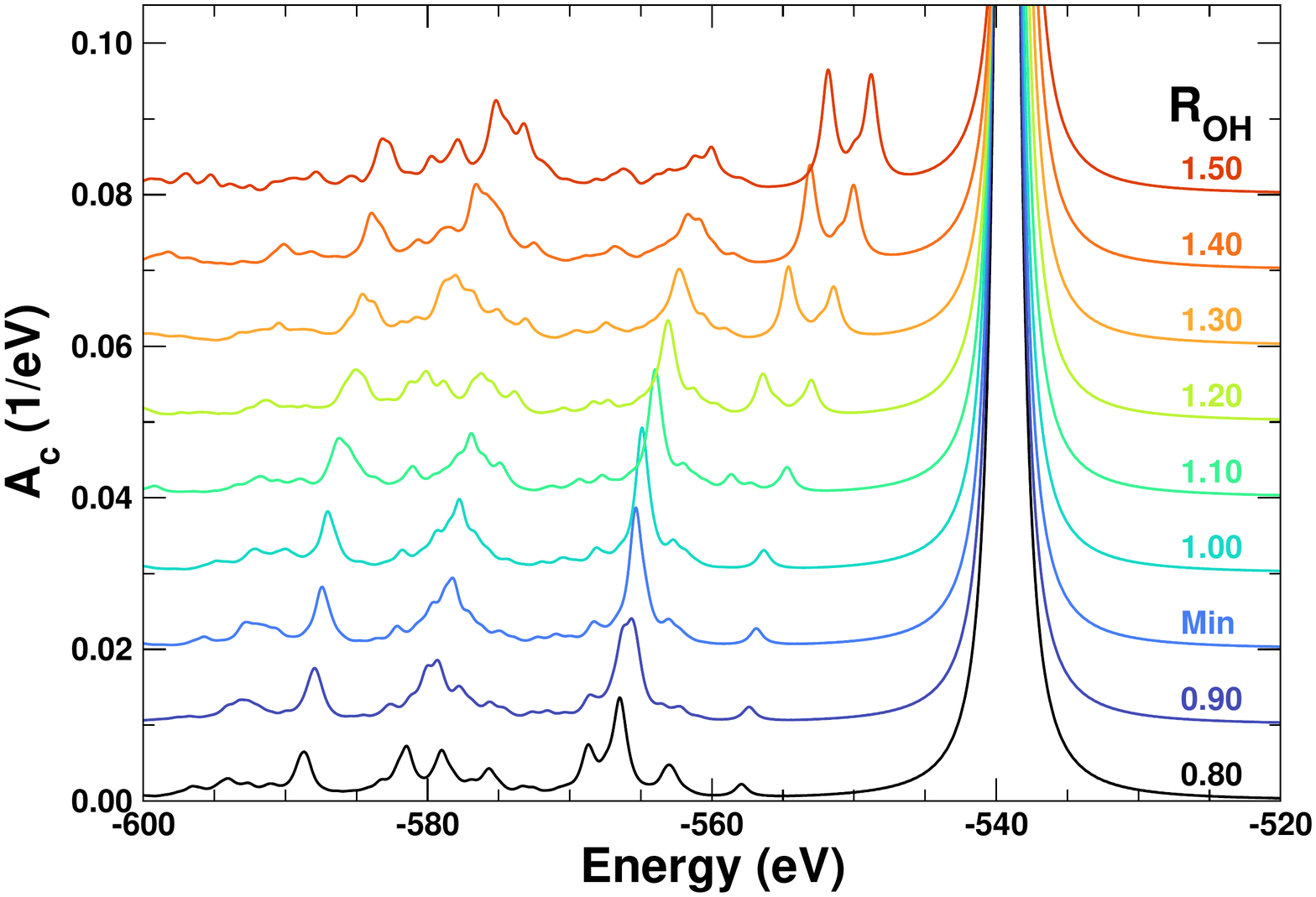}
	\includegraphics[clip,trim=0.5cm 1.0cm 0.5cm 2.8cm,
	width=0.45\textwidth]{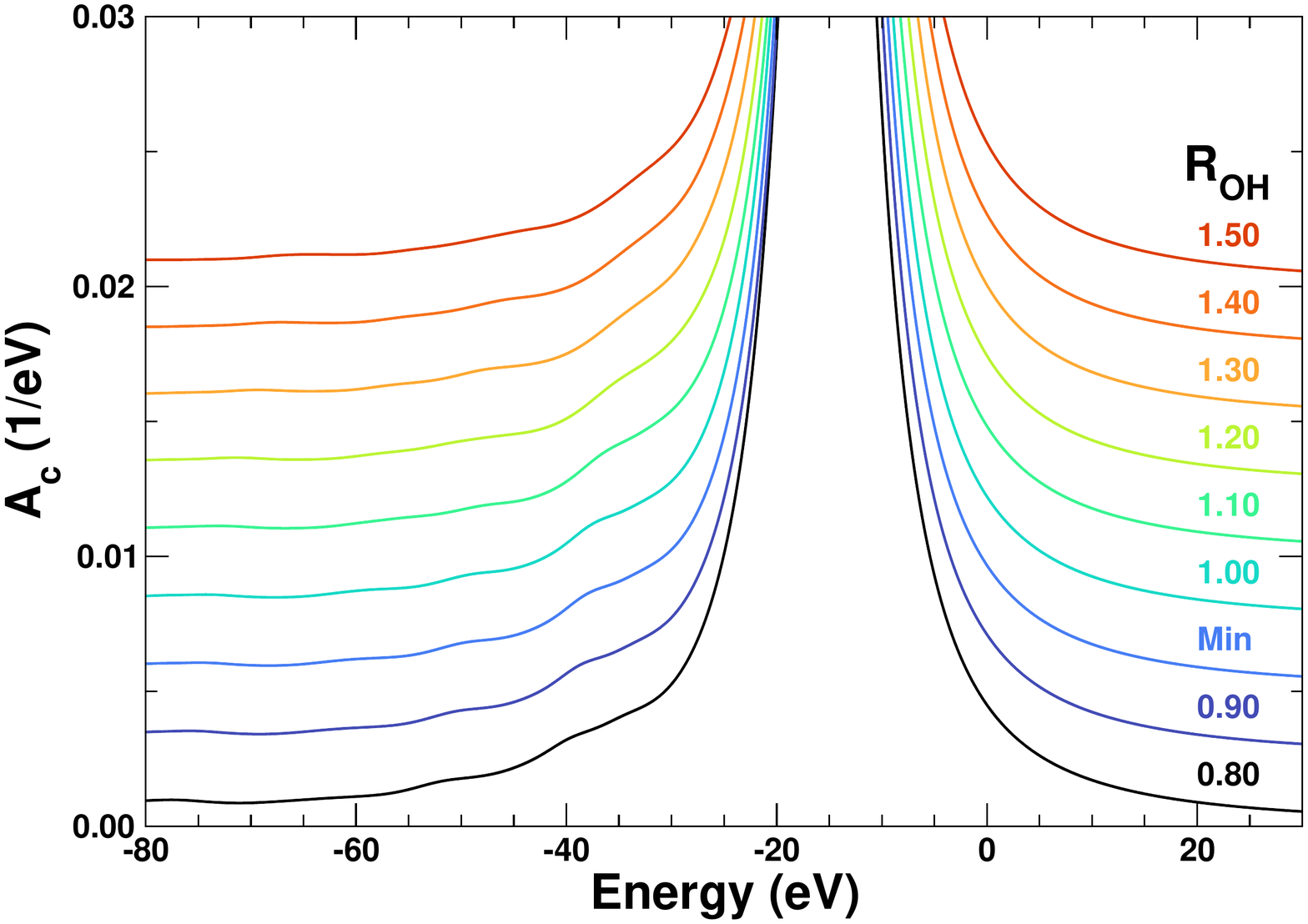}
	\includegraphics[clip,trim=0.5cm 1.0cm 0.5cm 2.8cm,
	width=0.45\textwidth]{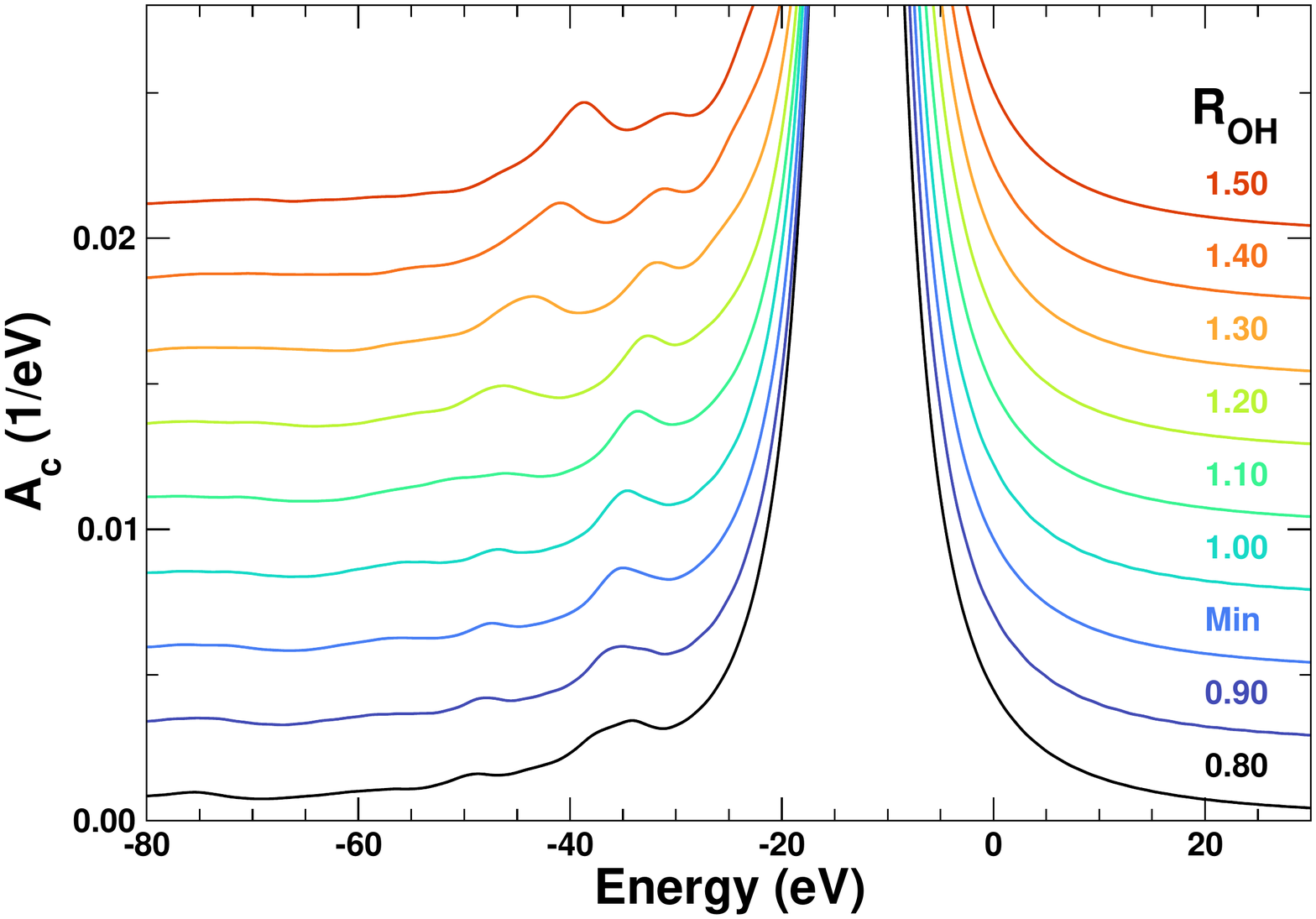}
	\caption{Quasi-particle peak and satellite region of the spectral function, as a function of OH bond length (R$_\mathrm{OH}$) for the core $^2A_1$ (top), HOMO-1 $1^2A_1$ (middle), and HOMO $1^2B_1$ (bottom) states of H$_2$O, calculated with the RT-EOM-CCSD method using the aug-cc-pVTZ basis set.}
\label{fig:h2oSPF}
\end{figure}

Fig.~\ref{fig:h2oIP} (bottom) shows the ionization energies and QP strengths for the HOMO $1^2B_1$ and HOMO-1 $1^2A_1$ states of H$_2$O, computed with the RT-EOM-CCSD method and the aug-cc-pVTZ basis set. For these states the RT-EOM-CCSD method gives good results, with errors compared to experiment of only 0.02 eV and 0.03 eV, respectively. Both $1^2B_1$ and $1^2A_1$ states are sensitive to the bond stretching, with a decrease in ionization energy of ~1 eV along the bond dissociation. The ionization energies have very similar behavior over the whole range, yet, for these states the QP strength shows very different behavior. While the QP strength of the $1^2B_1$ state decreases monotonically $\sim$6\%, the one for the $1^2A_1$ varies only $\sim$2\% but goes through a minimum at 1.1 \AA. This can be clearly seen in the middle panel of Fig. \ref{fig:h2oSPF}, that shows that the spectral function of the $1^2A_1$ state is nearly featureless in the satellite region, in accordance with the little weight transfer from the QP. For the $1^2B_1$ state (Fig.~\ref{fig:h2oSPF}, bottom), the $\sim$6\% weight transfer is manifested in a the small increase in satellite weight in the -50 to -30 eV region.

\begin{figure}
	\includegraphics[clip,trim=0.5cm 1.0cm 0.5cm 2.8cm,
	width=0.50\textwidth]{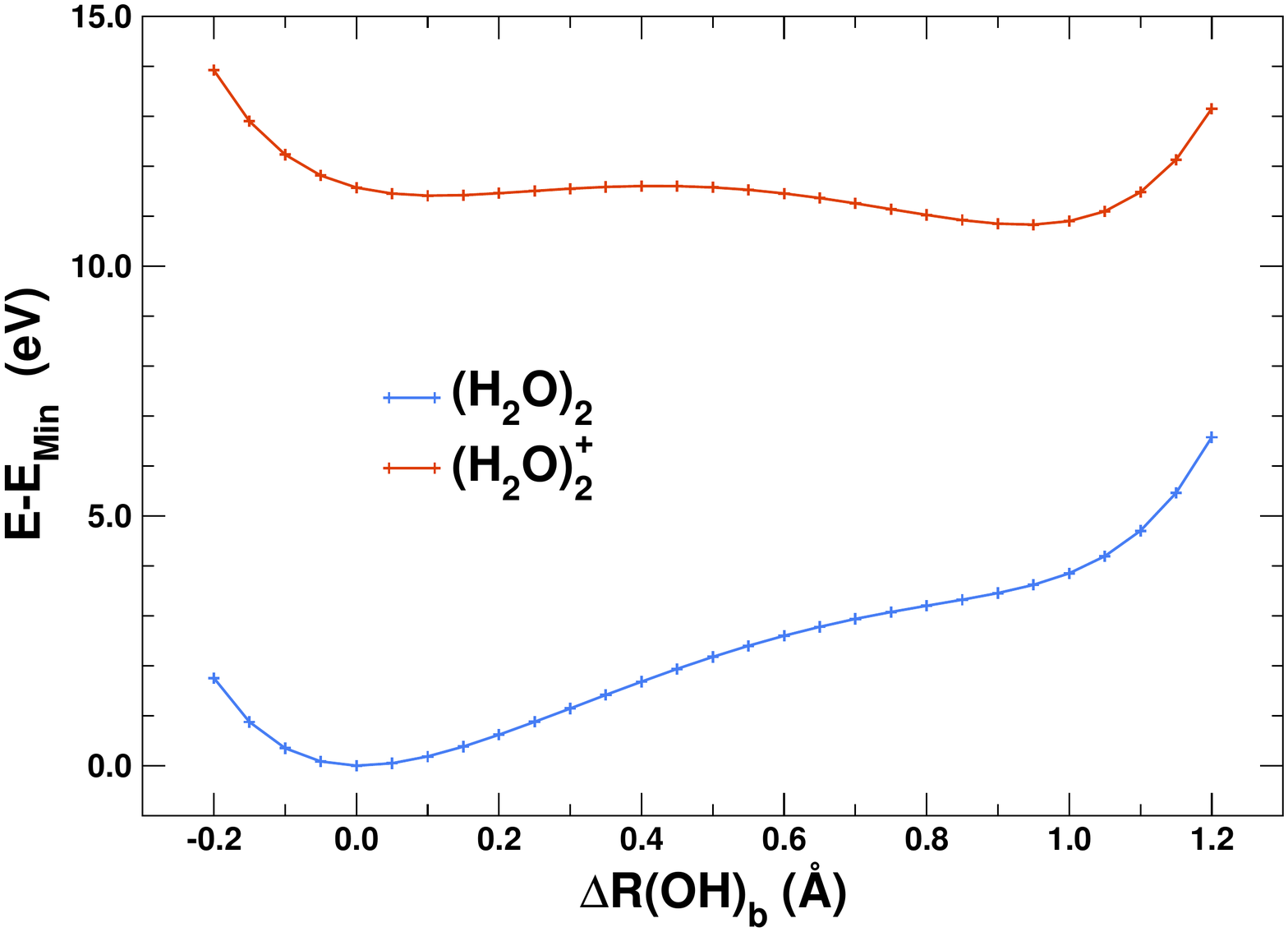}
	\includegraphics[clip,trim=0.5cm 1.0cm 0.5cm 2.8cm,
	width=0.50\textwidth]{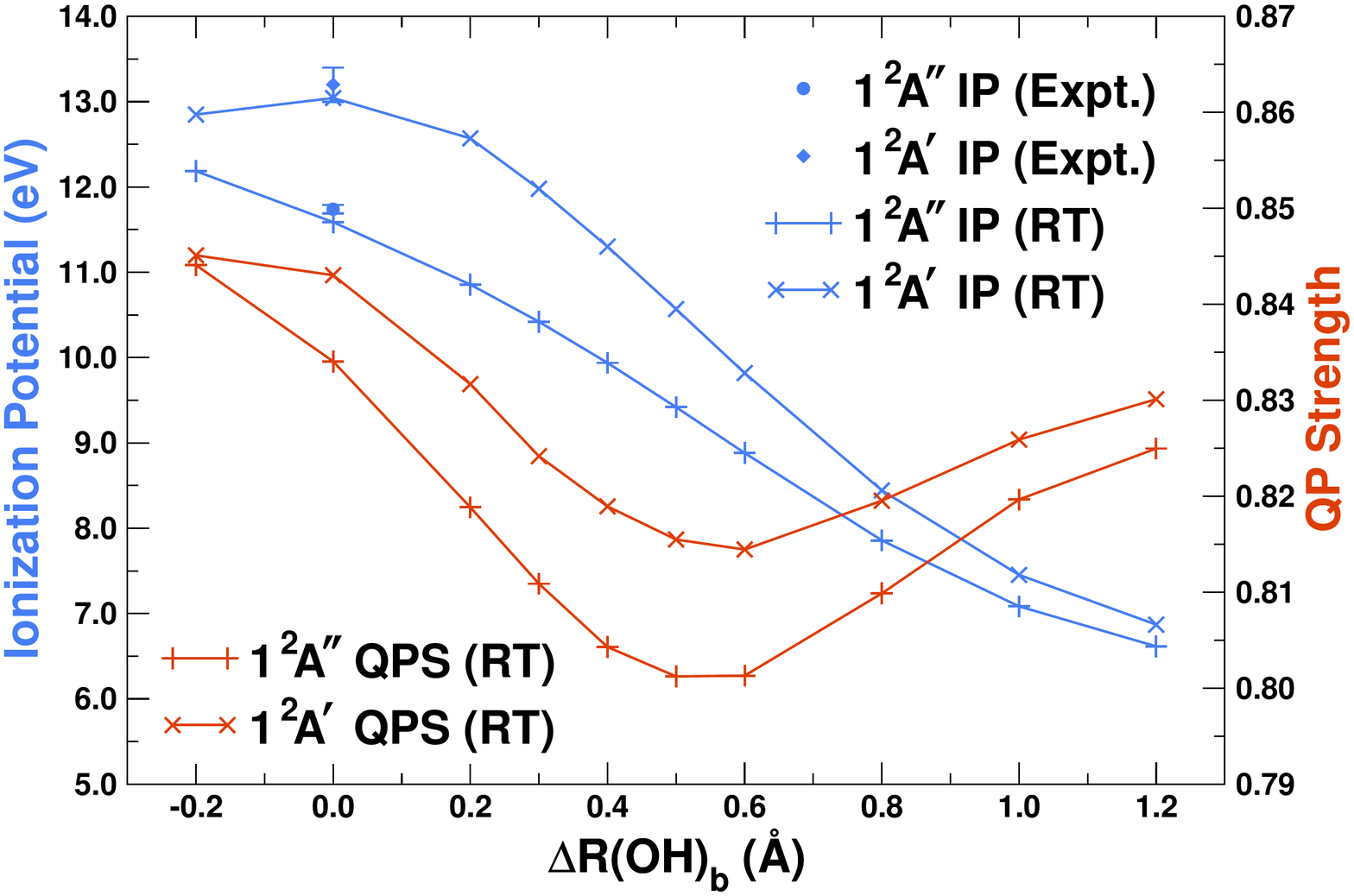}
	\caption{Top: Potential energy surface for neutral and ionized (H$_2$O)$_2$ as a function of proton transfer coordinate $\Delta$R(OH)$_\mathrm{b}$ calculated at the CCSD level. Bottom: Ionization potentials (IP) and quasi-particle strengths (QPS) as a function of proton transfer coordinate $\Delta$R(OH)$_\mathrm{b}$ for the HOMO $1^2A''$ and HOMO-1 $1^2A'$ states of (H$_2$O)$_2$, calculated with the RT-EOM-CCSD method (RT). All calculations used the aug-cc-pVDZ basis set. The blue circles and diamonds indicate the experimental (Expt.) ionization energies at equilibrium.\cite{tomoda1982photoelectron,stephen_leone_water_dimer}}
\label{fig:h2o2IP}
\end{figure}

\begin{figure}[t]
	\includegraphics[clip,trim=0.5cm 1.0cm 0.5cm 2.8cm,
	width=0.50\textwidth]{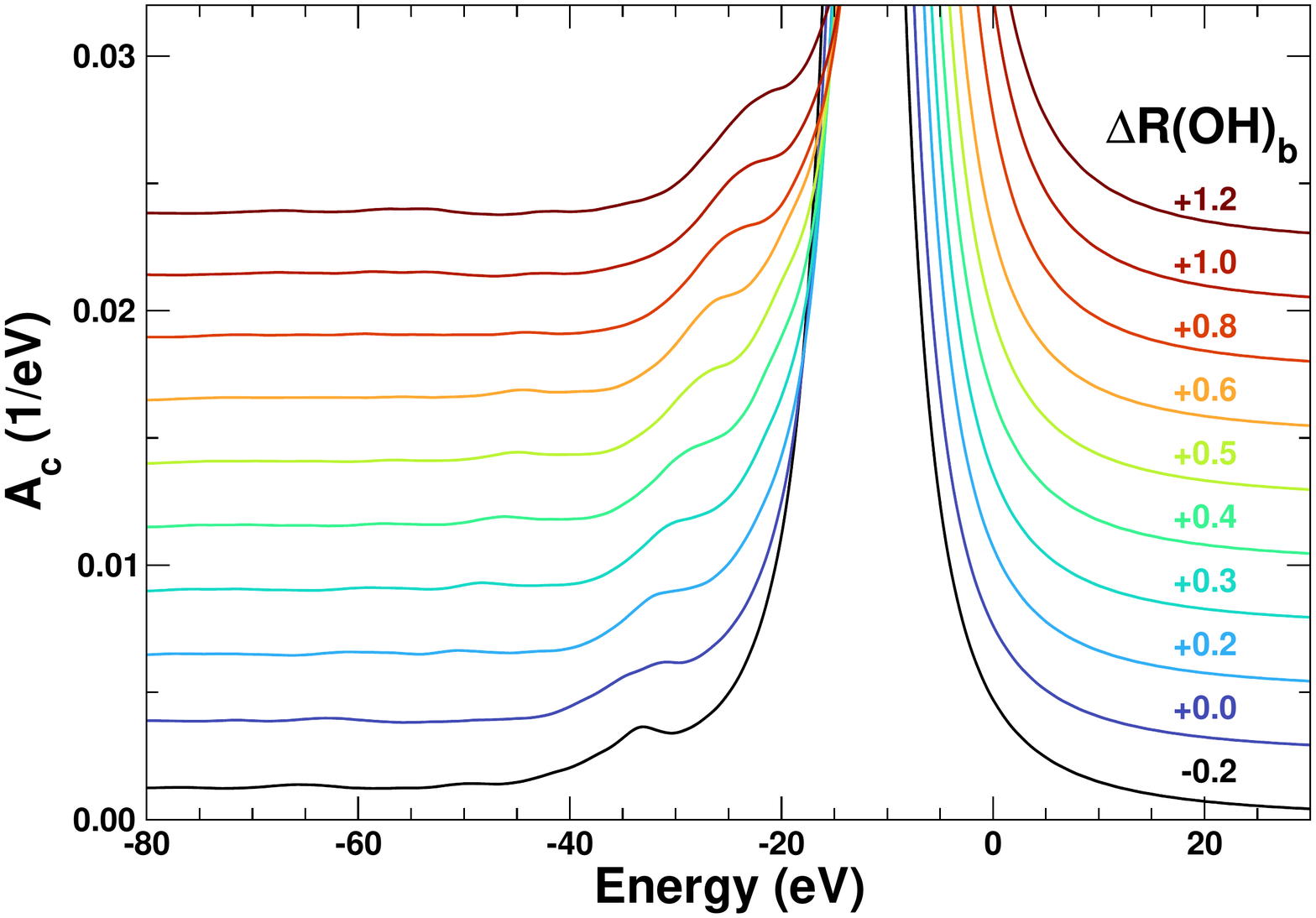}
	\includegraphics[clip,trim=0.5cm 1.0cm 0.5cm 2.8cm,
	width=0.50\textwidth]{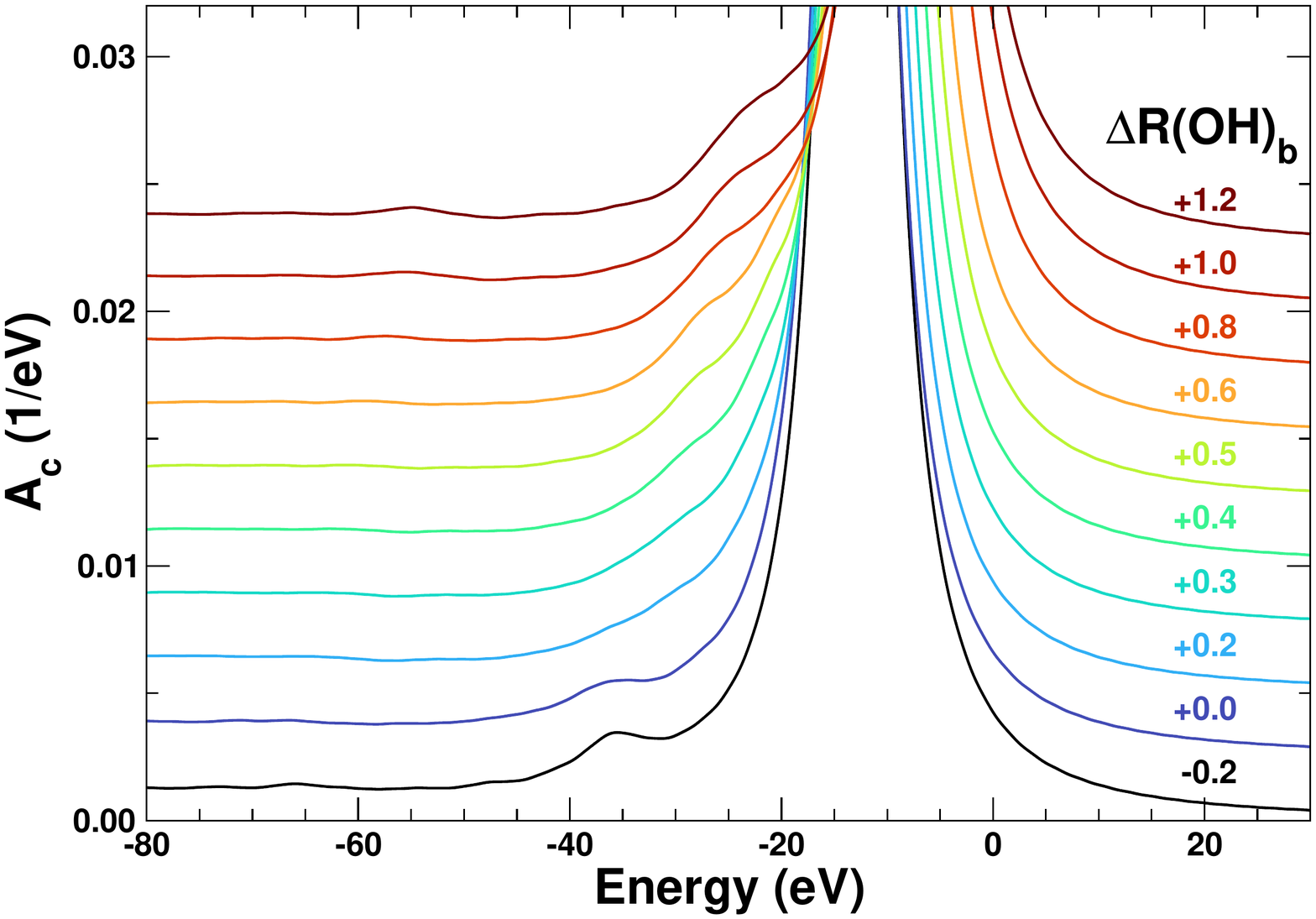}
	\caption{Quasi-particle peak and satellite region of the spectral function as a function of proton transfer coordinate $\Delta$R(OH)$_\mathrm{b}$ for the HOMO $1^2A''$ (top) and HOMO-1 $1^2A'$ (bottom) states of (H$_2$O)$_2$, calculated with the RT-EOM-CCSD method. All calculations used the aug-cc-pVDZ basis set.}
\label{fig:h2o2SPF}
\end{figure}

Fig.~\ref{fig:h2o2IP} (bottom) shows results for the ionization of the HOMO $1^2A''$ and HOMO-1 $1^2A'$ states of (H$_2$O)$_2$ as a function of proton transfer coordinate. For reference, the top panel shows the potential energy surfaces for the ground and first ionized ($1^2A''$) states computed at the standard CCSD level. Unlike the monomer case, which shows little variation as a function of coordinate, the two lowest ionized states of the dimer vary significantly. The predicted decrease in IP associated with the proton transfer is expected given the formation of an (OH)$^-$(H$_3$O)$^+$ cluster. The IP of this proton-transferred system should be dominated by the ionization of the (OH)$^-$ fragment, which has a low IP despite being``solvated'' by the (H$_3$O)$^+$ fragment.\cite{winter2006electron} Moreover, the first two ionized states of (OH)$^-$(H$_3$O)$^+$ are expected to become nearly degenerate given that the HOMO state of isolated OH$^-$ is doubly degenerate (1$\pi$). The CCSD PES also predicts an adiabatic ionization energy from (H$_2$O)$_2$ to disproportionated ion (OH)(H$_3$O)$^+$ of 10.83 eV, in reasonable agreement with other estimated values of 10.64 eV.\cite{barnett1995pathways} The QP strength (Fig.~\ref{fig:h2o2IP}) shows a small ($\sim$4\%) variation over the proton transfer range for the $1^2A''$ state, and goes through a minimum in the region 0.4-0.6 \AA. The $1^2A'$ state shows very similar behavior. The QP strength minima coincide with the transition state region in the ionized state potential energy surface and result from an increase in many-body electron correlation in the bond breaking/reforming region, which manifests itself in the spectral function as a transfer of weight to the satellites (Fig.~\ref{fig:h2oSPF}). Beside the QP shift as a function of proton transfer coordinate, which actually results in an overall shift, the spectral function shows very little change, as seen from Fig.~\ref{fig:h2o2SPF}. 

\begin{figure}[t]
    \includegraphics[clip,trim=0.5cm 1.0cm 0.5cm 2.8cm,
	width=0.50\textwidth]{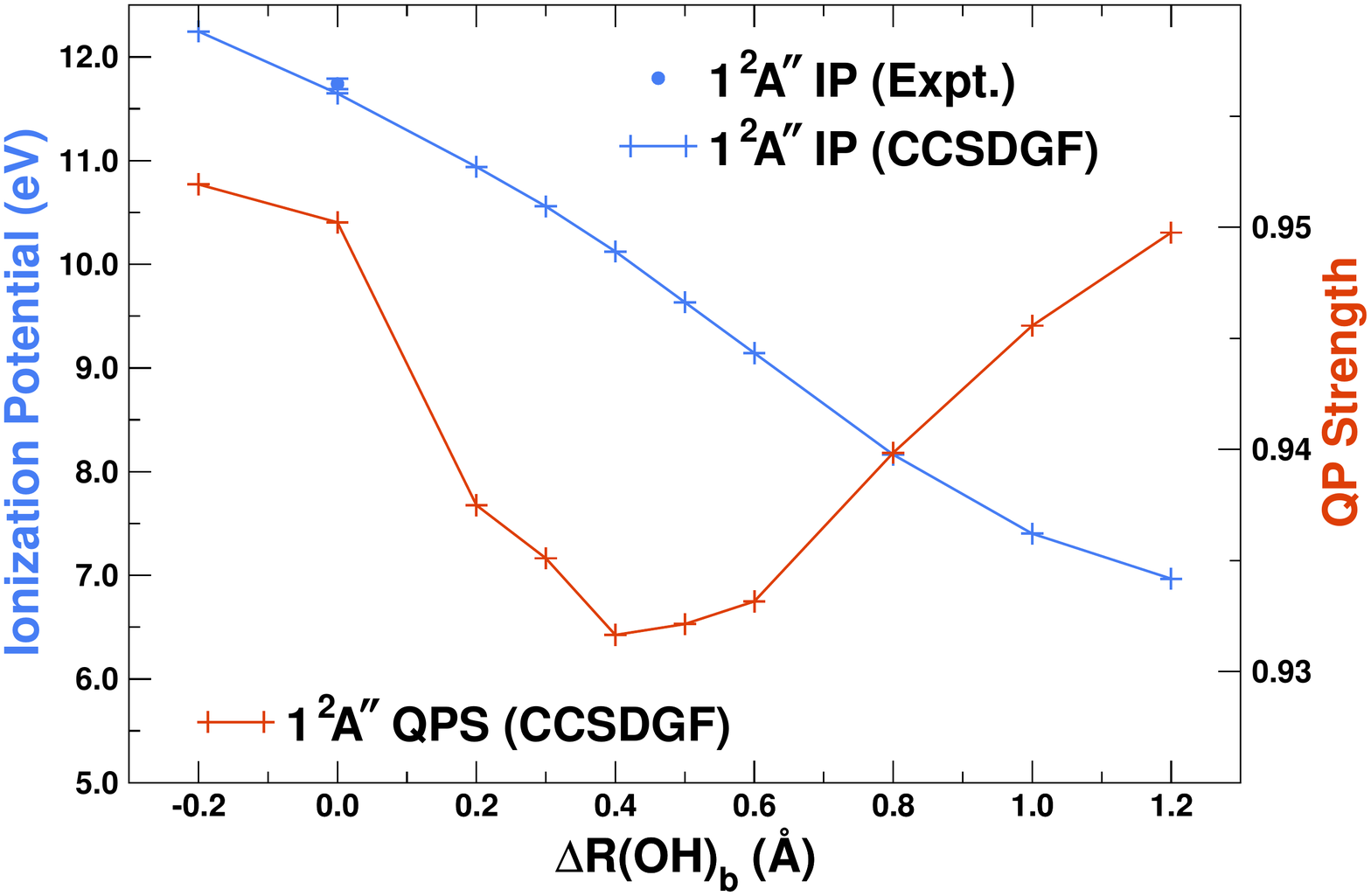}
	\includegraphics[clip,trim=0.5cm 1.0cm 0.5cm 2.8cm,
	width=0.50\textwidth]{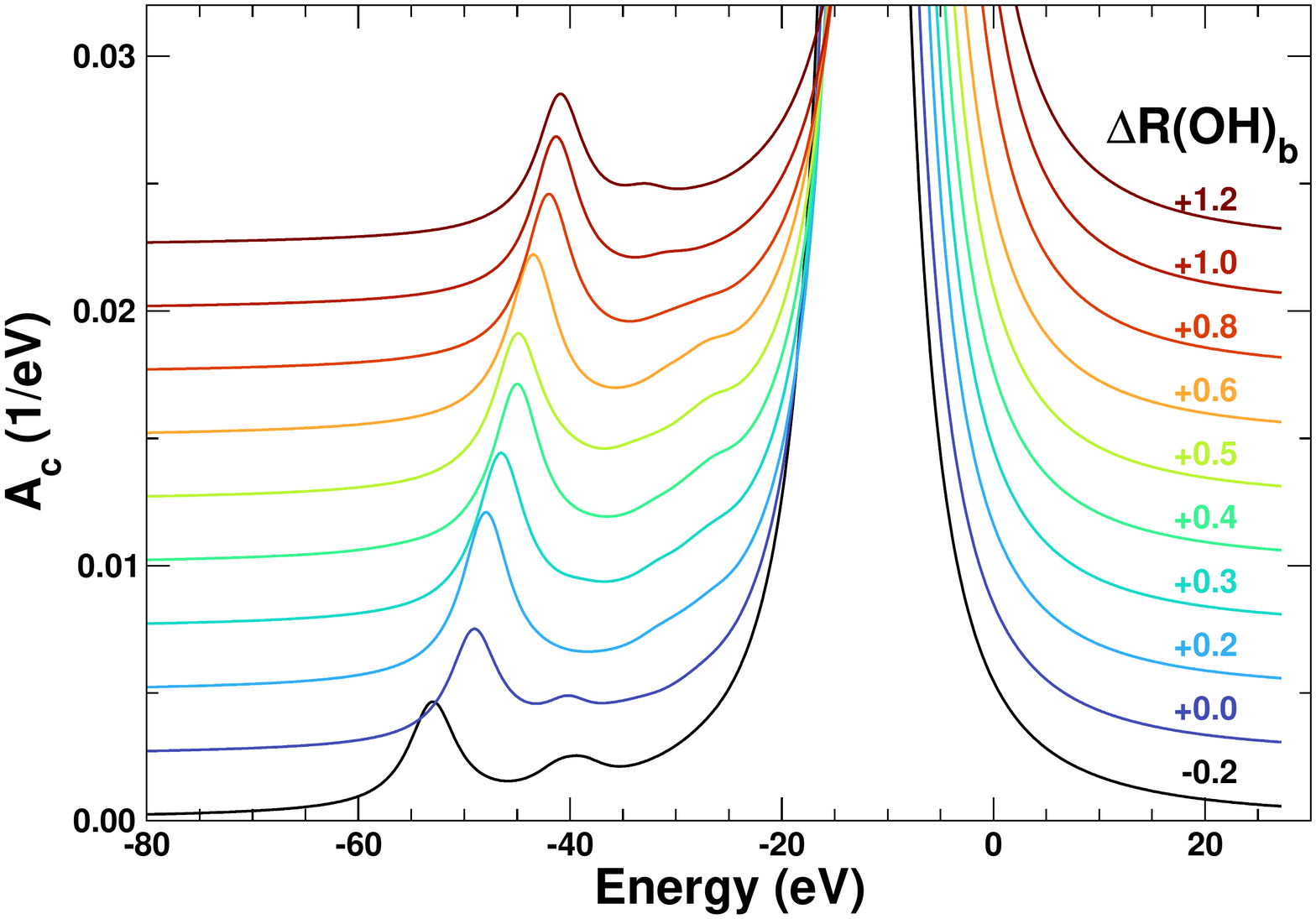}
	\caption{Ionization potential (IP) and quasi-particle strength (QPS) (top) and quasi-particle peak and satellite region of the spectral function (bottom) as a function of proton transfer coordinate $\Delta$R(OH)$_\mathrm{b}$ for the HOMO $1^2A''$ state of (H$_2$O)$_2$, calculated with the CCSDGF method. All calculations used the aug-cc-pVDZ basis set. The blue circle indicates the experimental (Expt.) ionization energy at equilibrium.\cite{stephen_leone_water_dimer}}
\label{fig:h2o2GFCCSD}
\end{figure}

We are interested in the change in satellite features as a function of proton transfer coordinate, $\Delta$R(OH)$_\mathrm{b}$. From Fig.~\ref{fig:h2o2SPF} a shoulder satellite peak can be observed on the l.h.s. of the main QP peak at all the studied $\Delta$R(OH)$_\mathrm{b}$'s, and the satellite peak position red-shifts as the $\Delta$R(OH)$_\mathrm{b}$ gets larger. This observation is also confirmed by independent static CCSDGF calculations using the same basis set for the same (H$_2$O)$_2$ systems at the same $\Delta$R(OH)$_\mathrm{b}$'s and over the same energy regime, and the corresponding results are shown in Fig.~\ref{fig:h2o2GFCCSD}. The only difference between Figs.~\ref{fig:h2o2SPF} and \ref{fig:h2o2GFCCSD} is that in the CCSDGF results the satellites are more separated from the QP and more refined satellite structure can be observed. At least two shoulder satellites are observed on the l.h.s. of the QP peak, while in the RT-EOM-CCSD results the refined satellite structure are embedded in the background of the QP peak. These peaks would likely be better resolved if with longer RT-EOM-CCSD calculations. Further wave function analysis of the satellites features show a $\pi\rightarrow \sigma*$ transition (e.g. from HOMO-1 to LUMO) along with the main ionization process from the HOMO. However, it is worth noting that the satellites predicted by CCSDGF are usually only qualitative, and systematic improvement towards quantitative evaluation would require including higher-order excitation in the CCGF framework as already demonstrated in the previous CCGF studies (see for example Ref. \citenum{peng2018green}). On the other hand, previous studies for small and medium-size molecules (e.g. H$_2$O, NH$_3$, CH$_4$, etc.)\cite{RVKNKP,vila2021eom,vila2020real,vila2022real} suggested that RT-EOM-CCSD is capable of providing better predictions for the satellite positions attributed to its explicit consideration of the time evolution of the correlation in $N-1$ electron state. Given the fact that only singles and doubles are included in both CCSDGF and RT-EOM-CCSD calculations, and the satellite peaks are close to each other, we conjecture that the satellite peak on the l.h.s. of the main QP peak also features a $\pi\rightarrow \sigma*$ transition along with the main ionization. It is worth mentioning that the satellite assignment in RT-EOM-CCSD method is challenging, and new methods in this regard are now under intensive development by the same authors.

\subsection{Core binding energies}

Fig.~\ref{fig:h2oIP} (top) shows the ionization energy and QP strength for the core state of the water monomer, computed with the RT-EOM-CCSD method and the Sapporo-TZP basis set. Unlike with the valence states, the core ionization energy is nearly insensitive to the deformation of the bond, with a change of only $\sim$0.7 eV. At the experimental bond length the agreement with the experimental core ionization energy error of about 0.5 eV is larger than for the valence states yet reasonable. On the other hand, the core ionization QP strength is much more sensitive to bond length than for the valence states, decreasing by about 10\% over the same range. This implies a significant transfer of weight from the QP to the satellite region. This transfer can be seen in Fig.~\ref{fig:h2oSPF} (top) which focuses on the satellite region of the spectral function as a function of bond length. Unlike the QP, which barely changes position, the satellite region 60 eV below it shows remarkable changes in weight distribution. As the bond dissociates, the main feature, 26 eV below the QP, becomes less pronounced and is replaced by a pair of features about 9 and 12 eV below the QP.

\section{Conclusions}
\label{conclusion}

In the present studies, we have compared the performance of several high-accuracy coupled-cluster formulations in calculating the ionization potentials and spectral functions of the water monomer and dimer for various geometries corresponding to bond-breaking and proton transfer processes. We established that the ionization potentials calculated with the CCSDGF, IP-EOMCCSD, and RT-EOM-CCSD methods reproduce those calculated with the CCSD(T) and CCSDT approaches for those systems. When compared to experiment, these methods provide accurate results, with average errors on the order of 0.1 eV. The direct comparison and crosscheck between the CCSDGF and RT-EOM-CCSD results also provides insights into the possible many-body excitations responsible for the satellite region above the quasiparticle peak. The peak assignment can be further improved by including higher-order excitation in the CCGF and RT-EOM-CC frameworks, and new methods are currently being explored. We find that the satellite region of the water monomer is particularly sensitive to bond breaking.
We have also demonstrated that the CCSD(T) formalism can almost perfectly reproduce the CCSDT ionization potentials for cases where CCSDT calculations were possible for neutral and charged systems. This prompted us to study the effect of basis set quality on the accuracy of the CCSD(T) ionization potentials. We find that for the water dimer, the aug-cc-pVTZ and aug-cc-pV5Z basis sets provides results converged to 0.1 eV and 0.01 eV, respectively. The largest CCSD(T)/aug-cc-pV6Z calculations for the (H$_2$O)$_2$ and (H$_2$O)$_2^+$ systems involved 1120 orbitals. While both CCSD(T) IPs obtained with the cc-pV6Z and aug-cc-pV6Z basis are almost identical, the discrepancy between calculated IPs obtained with the  smallest and the largest basis sets of a given type is significantly larger for the cc-pVXZ series and amounts to 0.89 eV, showing that the augmented basis set provide faster convergence.
Finally, we also reported results with the fully 4-component relativistic IP-EOMCCSD method for the lowest three states of the water monomer and found that its predictions are precise in comparison to the available experimental values. However, we found the effects of relativity to be almost negligible for the valence ionization potential values.   

\section{Acknowledgements}
This work was supported by the Computational Chemical Sciences Program of the U.S. Department of Energy, Office of Science, BES, Chemical Sciences, Geosciences and Biosciences Division in the Center for Scalable and Predictive methods for Excitations and Correlated phenomena (SPEC) at PNNL, with computational support from NERSC, a DOE Office of Science User Facility, under contract no. DE-AC02-05CH11231. B.P. also acknowledges support from the Laboratory Directed Research and Development (LDRD) Program at PNNL.

\section*{AUTHOR DECLARATIONS}
\subsection*{Conflict of Interest}
The authors have no conflicts of interest to declare.

\section*{DATA AVAILABILITY}
The data that support the findings of this study are available from the corresponding authors
upon reasonable request.

\section*{REFERENCES}
\bibliographystyle{h-physrev}


\end{document}